\documentstyle[11pt,newpasp,twoside]{article}
\def\edcomment#1{\iffalse\marginpar{\raggedright\sl#1\/}\else\relax\fi}
\reversemarginpar

\def\spose#1{\hbox to 0pt{#1\hss}}
\def\rsun{R_{\odot}}
\def\msun{M_{\odot}}
\def\rhosun{\rho_{\odot}}
\def\rstar{R_{*}}
\def\mstar{M_{*}}
\def\rplanet{R_{p}}

\def\mp{M_p}
\def\rp{R_p}
\def\sqdepth{\sqrt{\Delta F}}

\def\rhostar{\rho_*}

\begin{document}
\title{Extrasolar Planet Transit Light Curves and a Method to Select the 
Best Planet Candidates for Mass Follow-up}

\author{S.\ Seager}
\affil{Institute for Advanced Study, Einstein
Drive, Princeton, NJ 08540 and The Carnegie Institution of Washington,
Dept. of Terrestrial Magnetism, 5241 Broad Branch Rd. NW, Washington,
DC 20015 }
\author{G.\ Mall\'en-Ornelas}
\affil{Princeton University Observatory, Peyton Hall, Princeton, NJ 08544 and
P.\ Universidad Cat\'olica de Chile, Casilla 306, Santiago 22, Chile}

\begin{abstract}
A unique analytical solution of planet and star parameters can be
derived from an extrasolar planet transit light curve under a number
of assumptions.  This analytical solution can be used to choose the
best planet transit candidates for radial velocity follow-up
measurements, with or without a known spectral type. In practice, high
photometric precision ($<$ 0.005 mag) and high time sampling ($<$ 5
minutes) are needed for this method.  See Seager \& Mall\'en-Ornelas
(2002) for full details.
\end{abstract}

\section{Assumptions}
The following assumptions and conditions are necessary for a light
curve to yield a unique solution of planet and star parameters:\\
\noindent $\bullet$ The planet orbit is circular (valid for 
tidally-circularized extrasolar planets);\\
$\bullet$ $\mp$ $\ll$ $\mstar$ and the companion is dark compared 
to the central star;\\
$\bullet$ The stellar mass-radius relation is known;\\
$\bullet$ The light comes from a single star, rather than from 2 
or more blended stars;\\
$\bullet$ The eclipses have flat bottoms. This implies 
that the companion is fully superimposed on the central star's disk
and requires that the data is in a band pass where limb 
darkening is negligible;\\
$\bullet$ The period can be derived from the light curve 
(e.g., the two observed eclipses are consecutive).\\
In this article  $M$ is mass, $R$ is radius, $\rho$ is 
density, $P$ is period, $a$ is orbital semi-major axis, $i$ is the orbital inclination, and $G$ is the Gravitational constant. Where required
the subscript $p$ is for planet, $*$ for stellar, and $\odot$ for
solar.

\section{The Simplified Equations}
Five equations are used to uniquely solve for $\mstar$, $\rstar$,
$a$, $i$, and $\rp$. The simplified equations presented
below require the additional assumption that $\rstar \ll a$.

\begin{center}
Transit depth
\begin{equation}
\label{eq:depthorig} 
\Delta F \equiv \frac{F_{no \hspace{0.04in}
transit} - F_{transit}}{F_{no \hspace{0.04in} transit}} =
\left(\frac{\rplanet}{\rstar}\right)^2.
\end{equation}
Total transit duration
\begin{equation}
\label{eq:length} \label{eq:durationapprox} t_T = \frac{P \rstar
}{\pi a}\sqrt{\left(1 + \frac{\rp}{\rstar}\right)^2 -
\left(\frac{a}{R_*} \cos i\right)^2}.
\end{equation}
Transit shape ($t_F=$ flat part of transit and $t_T=$ total transit duration)
\begin{equation}
\label{eq:shapeapprox} \left(\frac{t_F}{t_T}\right)^2 = \frac{ \left( 1
- \frac{R_p}{R_*}\right)^2 - \left(\frac{a}{R_*} \cos i\right)^2}
{\left( 1 + \frac{R_p}{R_*}\right)^2 - \left(\frac{a}{R_*} \cos
i\right)^2}.
\end{equation}
Kepler's Third Law
\begin{equation}
\label{eq:Keplerorig}
P^2 = \frac{4 \pi^2 a^3}{G\mstar}.
\end{equation}
Stellar mass-radius relation
\begin{equation}
\label{eq:MRorig}
\rstar = k \mstar^x.
\end{equation}
\end{center}

Here $k$ is a constant coefficient for each stellar sequence (main
sequence, giants, etc.) and $x$ describes the power law of the
sequence (e.g., $k=1$ and $x \simeq 0.8$ for F--K main sequence stars
(Cox 2000)).  Note that Kepler's Third Law and the stellar mass-radius
relation set a physical scale to two disks passing in front of each
other. This breaks the geometrical degeneracy and allows a unique
solution.

\section{The Simplified Solution}
The five parameters $\mstar$, $\rstar$, $a$, $i$, and $\rp$ can be
solved for uniquely from the above five equations.  Moreover, 
the impact parameter $b\equiv a \cos i / \rstar$ and stellar density
$\rhostar$ can be solved for uniquely without the stellar
mass-radius relation.
\begin{center}
\begin{equation}
 \label{eq:bapprox}
  b = \left[ \frac{ (1 - \sqdepth)^2 -
 \left(\frac{t_F}{t_{T}}\right)^2 (1 + \sqdepth)^2}{1 -
 \left(\frac{t_F}{t_T}\right)^2} \right]^{1/2}.
\end{equation}
\vspace{0.05in}
\begin{equation}
\label{eq:rhoapprox} 
\frac{\rhostar}{\rhosun} = \frac{32}{G \pi} P
\frac{ \Delta F^{3/4}}{\left({t_T}^2 - t_F^2\right)^{3/2}}.
\end{equation}
\begin{equation}
 \label{eq:morig}
 \frac{\mstar}{\msun} = \left[ k^3 \frac{\rhostar}{\rhosun} \right]^{\frac{1}{1-3x}}.
\end{equation}
\vspace{0.05in} 
\begin{equation}
\label{eq:rorig}
\frac{\rstar}{\rsun} = k \left(\frac{\mstar}{\msun}\right) ^x =
\left[k^{1/x} \frac{\rhostar}{\rhosun}\right]^{\frac{x}{(1-3x)}}.
\end{equation}
\vspace{0.05in} 
\begin{equation}
\label{eq:aorig}
 a = \left[ \frac{P^2 G \mstar} {4
\pi^2}\right]^{1/3}.
\end{equation}
\vspace{0.05in} 
\begin{equation}
\label{eq:iorig}
 i = \cos^{-1}\left(b \frac{\rstar}{ a}\right).
\end{equation}
\vspace{0.05in} 
\begin{equation}
\label{eq:rporig}
 \frac {\rp}{\rsun} = \frac{\rstar}{\rsun} \sqdepth = \left[k^{1/x} \frac{\rhostar}{\rhosun}\right]^{\frac{x}{(1-3x)}} \sqdepth.
\end{equation}
\end{center}

\begin{figure}[ht]
\plotfiddle{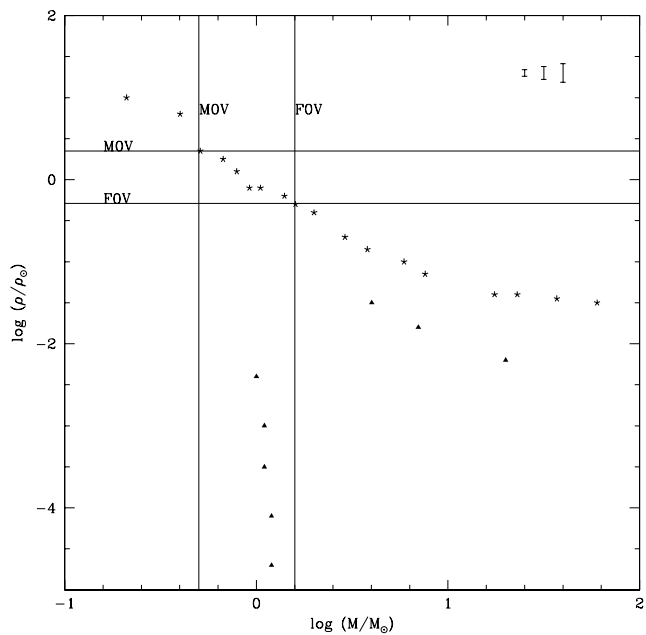}{3in}{0}{45}{45}{-140}{-80}
\caption{Stellar density $\rhostar$ vs. stellar mass $\mstar$ ($\mstar$ is used
as a proxy for stellar spectral type). See text for details. The box MOV to F0V shows the main sequence stars which are most appropriate for finding transiting planets. See Seager \& Mall\'en-Ornelas (2002) for a discussion of errors.}
\end{figure}

\section{Application}
The above analytical solution has many applications, all related to
selecting the best transit candidates for radial velocity mass
follow-up.  Here we only have room to describe one application; for 
others see Seager \& Mall\'en-Ornelas
(2002).

The stellar density $\rhostar$ can be uniquely determined from the
light curve alone without using the stellar mass-radius relation, as
seen from equation~(7).  
%
A measured $\rhostar$ can be used in three ways.  (1) From the light
curve alone a main sequence star and a giant star can be distinguished
because main sequence stars occupy a unique position in a $\rhostar$
vs. spectral type diagram (Figure~1). Hence a giant star with an
eclipsing stellar companion can be ruled out.  (2) From the light
curve and the stellar mass-radius relation $\rp$ can be estimated
(equation~(12)). Even for slightly evolved stars an upper limit on
$\rstar$ and hence $\rp$ can be derived. (3) A common false
positive planet transit can be ruled out by comparing $\rhostar$
derived from the light curve with $\rhostar$ derived from a spectral
type. If the two $\rhostar$ differ then something is amiss with the
assumptions in \S1. The common case is the situation where a
binary star system has its eclipse depth reduced to a planet-size
eclipse due to the light from a third, contaminating, star (Figure
2). The contaminating star may be a chance alignment of a foreground
or background star, or a third star as part of a triple star system.
For a real example of this ``blended star'' situation, see
Mall\'en-Ornelas et al. (2002).
\begin{figure}[h]
\plotfiddle{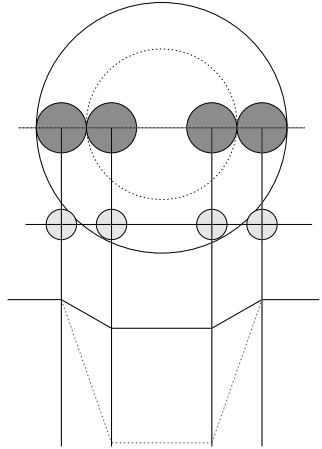}{2.2in}{0}{30}{30}{-70}{0}
\caption{A deep binary star eclipse (dotted line) can mimic
a planet transit (solid line) when extra light from a third, contaminating
star (not shown) is present.}
\end{figure}

\acknowledgements
This work was supported by the W. M. Keck Foundation
and the Carnegie Institution of Washington.

\end{document}